\documentclass[aps,prl,twocolumn,superscriptaddress,showpacs]{revtex4-1}

\usepackage{graphicx,bm,mathtools,color,amssymb}

\begin{document}
\title{Fully gapped $d$-wave superconductivity in CeCu$_2$Si$_2$}

\author{G. M. Pang}
\affiliation{Center for Correlated Matter and Department of Physics, Zhejiang University, Hangzhou 310058, China}
\author{M. Smidman}
\affiliation{Center for Correlated Matter and Department of Physics, Zhejiang University, Hangzhou 310058, China}
\author{J. L. Zhang}
\affiliation{Center for Correlated Matter and Department of Physics, Zhejiang University, Hangzhou 310058, China}
\author{L. Jiao}
\affiliation{Center for Correlated Matter and Department of Physics, Zhejiang University, Hangzhou 310058, China}
\author{Z. F. Weng}
\affiliation{Center for Correlated Matter and Department of Physics, Zhejiang University, Hangzhou 310058, China}
\author{E. M. Nica}
\affiliation{Department of Physics and Astronomy, Rice University, Houston, Texas 77005, USA}
\affiliation{Department of Physics and Astronomy and Quantum Materials Institute, University of British Columbia, Vancouver, B.C., V6T 1Z1, Canada}
\author{Y. Chen}
\affiliation{Center for Correlated Matter and Department of Physics, Zhejiang University, Hangzhou 310058, China}
\author{W. B. Jiang}
\affiliation{Center for Correlated Matter and Department of Physics, Zhejiang University, Hangzhou 310058, China}
\author{Y. J. Zhang}
\affiliation{Center for Correlated Matter and Department of Physics, Zhejiang University, Hangzhou 310058, China}
\author{W. Xie}
\affiliation{Center for Correlated Matter and Department of Physics, Zhejiang University, Hangzhou 310058, China}
\author{H. S. Jeevan}
\affiliation{Center for Correlated Matter and Department of Physics, Zhejiang University, Hangzhou 310058, China}
\affiliation{Experimentalphysik VI, Center for Electronic Correlations and Magnetism, University of Augsburg, 86159 Augsburg, Germany}
\author{H. Lee}
\affiliation{Center for Correlated Matter and Department of Physics, Zhejiang University, Hangzhou 310058, China}
\author{P. Gegenwart}
\affiliation{Experimental Physics VI, Center for Electronic Correlations and Magnetism, University of Augsburg, 86159 Augsburg, Germany}
\author{F. Steglich}
\affiliation{Center for Correlated Matter and Department of Physics, Zhejiang University, Hangzhou 310058, China}
\affiliation{Max Planck Institute for Chemical Physics of Solids, 01187 Dresden, Germany}
\author{Q. Si}
\affiliation{Department of Physics and Astronomy, Rice University, Houston, Texas 77005, USA}
\author{H. Q. Yuan}
\email{hqyuan@zju.edu.cn}
\affiliation{Center for Correlated Matter and Department of Physics, Zhejiang University, Hangzhou 310058, China}
\affiliation{Collaborative Innovation Center of Advanced Microstructures, Nanjing University, Nanjing 210093, China}

\begin{abstract}
The nature of the pairing symmetry of the first heavy fermion superconductor CeCu$_2$Si$_2$ has recently become the subject of controversy. While CeCu$_2$Si$_2$ was generally believed to be a $d$-wave superconductor, recent low temperature specific heat measurements  showed evidence for fully gapped superconductivity, contrary to the nodal behavior inferred from earlier results.  Here we report London penetration depth measurements, which also reveal fully gapped behavior at very low temperatures. To explain these seemingly conflicting results, we propose a fully gapped $d+d$ band-mixing pairing state for CeCu$_2$Si$_2$, which yields very good fits to both the superfluid density and specific heat, as well as accounting for a sign change of the superconducting order parameter, as previously concluded from inelastic neutron scattering results.
\end{abstract}
\maketitle

The structure of the superconducting order parameter has been frequently studied, due to its close relationship with the underlying pairing mechanism. While the conventional electron-phonon pairing mechanism typically leads to $s$-wave states with fully opened gaps and a constant sign over the Fermi surface \cite{Bardeen1957}, unconventional superconductors with different pairing mechanisms often form states with a sign changing order parameter \cite{Anderson-Science,ScalRMP}. 
For instance, cuprate and many Ce-based heavy fermion superconductors are generally believed to be $d$-wave superconductors, with nodal lines in the energy gap on the Fermi surface \cite{KotliarLiu,SigristUeda,CeTIn5Node1}. On the other hand, in the high temperature iron based superconductors, an  $s\pm$-state has been proposed, with a change of sign of the gap function between disconnected Fermi surface pockets, but the energy gap remains nodeless \cite{IronPairRev}. In this context,
the surprising recent discovery \cite{Kittaka2014,Yamashita2017,Takenaka2017} of evidence for fully gapped superconductivity   in the first heavy fermion superconductor CeCu$_2$Si$_2$  \cite{Steglich1979} requires further attention.

Superconductivity in CeCu$_2$Si$_2$  occurs in close proximity to magnetism. Samples
with either superconducting ($S$-type), antiferromagnetic ($A$-type) or competing phases ($A/S$-type) are obtained via slight tuning of the composition within the homogeneity range \cite{Steglich1996}. High pressure measurements of CeCu$_2$(Si$_{1-x}$Ge$_x$)$_2$ reveal two distinct superconducting domes, one centered around  an antiferromagnetic (AFM) instability at ambient/low pressure, and another near a valence instability at high pressure \cite{Yuan2003}. The close proximity of superconductivity to an AFM instability suggests that in CeCu$_2$Si$_2$ it is driven by the corresponding quantum criticality. Inelastic neutron scattering (INS) measurements clearly indicate that 
the Cooper pairing is associated with 
a damped propagating paramagnon mode at the incommensurate ordering wavevector ${\boldsymbol Q}_{\mathbf{AF}}$ 
of the spin-density wave (SDW) order nearby in the phase diagram \cite{Stockert2011}. 
The large intensity of the low energy spin excitation spectrum at ${\boldsymbol Q}_{\mathbf{AF}}$, 
which reveals a spin gap in the superconducting state, as well as a pronounced peak \textit{well inside} 
the superconducting gap $2\Delta\approx5k_BT_c$  \cite{Stockert2011,Fujiwara2008}, implies a sign change of the pairing function between the two regions of the Fermi surface spanned by ${\boldsymbol Q}_{\mathbf{AF}}$ \cite{SpinResTheor,SpinResTheor2}.  The absence of a coherence peak and the $\sim T^3$ temperature dependence of the spin-lattice relaxation rate [$1/T_1(T)$] in Cu-NQR measured above 100~mK, further suggested an unconventional superconducting order parameter with line nodes in the gap structure \cite{Ishidal1999,Fujiwara2008}. Angle resolved resistivity measurements at 40~mK indicate a four-fold modulation of the upper critical field $H_{c2}$, as expected for a $d$-wave gap with $d_{xy}$ symmetry \cite{Viyera2011}, while a sign change spanning ${\boldsymbol Q}_{\mathbf{AF}}$ is  compatible with $d_{x^2-y^2}$ pairing symmetry \cite{SpinResTheor}.  Therefore, CeCu$_2$Si$_2$ behaves as
an even-parity $d$-wave superconductor,
whose gap structure has yet to be determined.

However, a recent specific heat investigation reported exponential behavior of $C(T)/T$ at very low temperatures, suggesting fully gapped superconductivity in CeCu$_2$Si$_2$ \cite{Kittaka2014}. Following this work, scenarios of multiband superconductivity with a strong Pauli paramagnetic effect, loop-nodal $s_{\pm}$ superconductivity, and $s_{++}$ pairing with no sign change were proposed \cite{Tsutsumi2014,Ikeda2015,Yamashita2017,Takenaka2017}. Furthermore, scanning tunneling spectroscopy down to 20~mK also hints at a  multigap order parameter \cite{Mostafa2015}. Indeed electronic structure calculations reveal that multiple bands cross the Fermi level \cite{Kittaka2014,CCSBS}, and renormalized band structure calculations show that the dominant heavy band (with $m^*/m_e\approx500$) leads to Fermi surface sheets mainly consisting of warped cylinders along the $c$~axis \cite{CCSBS}. The 
aforementioned
discrepancies between the pairing symmetries deduced from  different measurements show that the superconducting order parameter of CeCu$_2$Si$_2$ is poorly understood. A particular puzzle is how to reconcile the fully gapped behavior with the previous evidence for a sign changing order parameter and nodal superconductivity. Here we probe the superconducting gap symmetry by measuring the temperature dependence of the London penetration depth, and propose a new scenario of a  fully gapped $d+d$ band-mixing pairing state, which reconciles all the
 seemingly contradictory results. 

\section*{Results}
\subsection*{Resistivity and specific heat}
The samples were characterized using resistivity and specific heat measurements, as shown for the $S$-type sample in  Fig.~1(a). The residual resistivity of the $S$-type sample in the normal state just above $T_c$ is $\rho_0~\approx~40~\mu\Omega$-cm, and a superconducting transition is observed, onsetting around 0.65~K and reaching zero resistivity at about 0.6~K. The transition width of $\approx0.05$~K is in line with recent reports \cite{Viyera2011}. The specific heat also shows a superconducting transition with $T_c~\approx~$0.64~K, similar to previous results  \cite{Kittaka2014}. The $A/S$-type sample [Fig.~1(b)] displays a superconducting transition, onsetting around 0.62~K, with a lower residual resistivity of  $\rho_0~\approx~12~\mu\Omega$-cm. The specific heat shows both an AFM transition at  $T_N~\approx~$0.7~K and a superconducting transition at  $T_c~\approx~$0.53~K.

\begin{figure}[t]
\centering
\includegraphics[width=.99\linewidth]{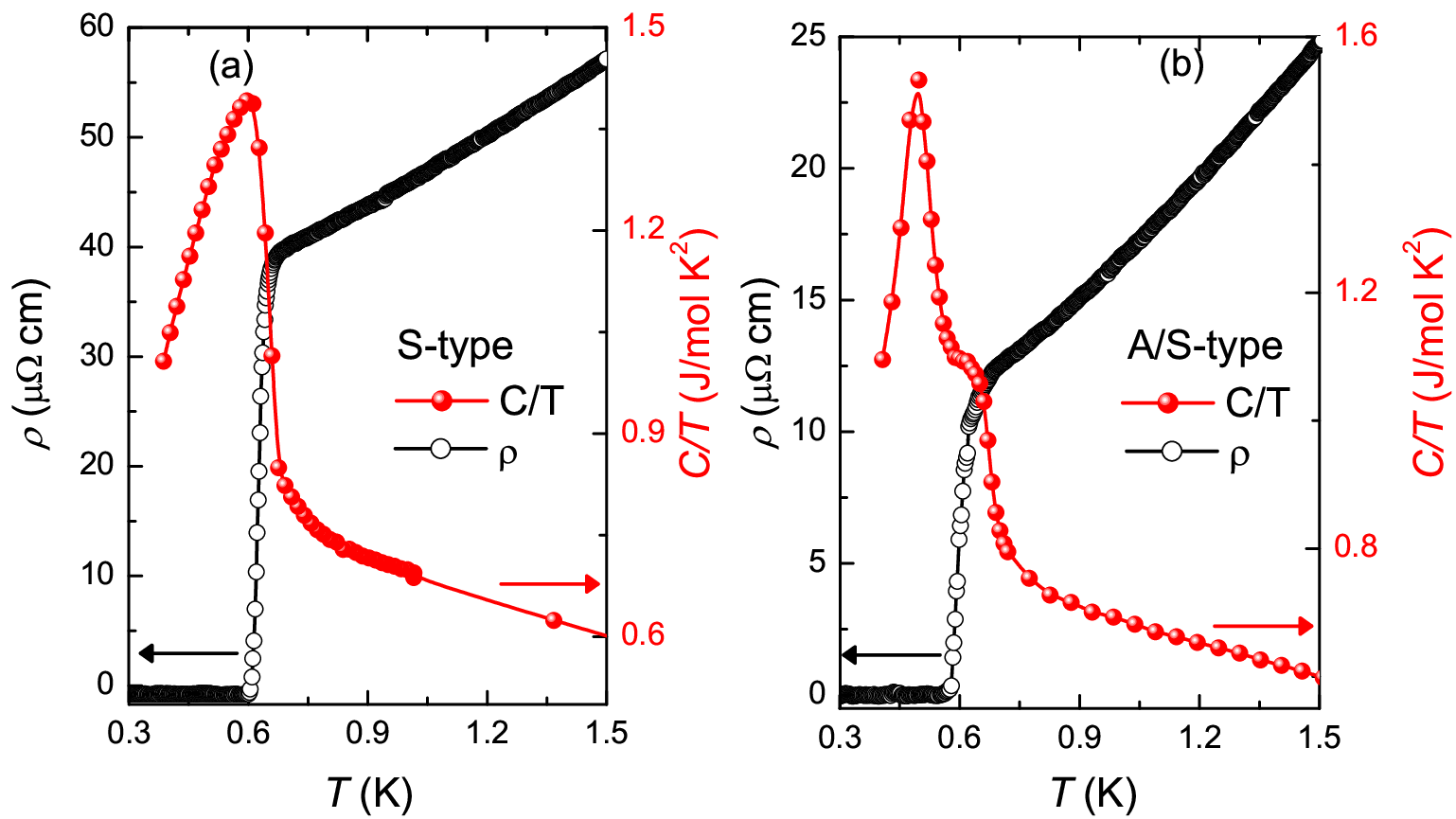}
\caption{ Specific heat as $C(T)/T$ and resistivity $\rho(T)$ of (a) $S$-type  and (b) $A/S$-type  CeCu$_2$Si$_2$. }
\label{CCSChar}
\end{figure}

\subsection*{Temperature dependence of the penetration depth}
Measurements of the change of the London penetration depth  $\Delta\lambda(T)~=~\lambda(T)-\lambda(0)$ for the  $S$-type sample are  displayed in Fig.~2(a). As shown in the inset, a sharp superconducting transition is clearly observed, with an onset at around 0.62~K. To probe the superconducting gap structure,  we analyzed the behavior of  $\Delta\lambda(T)$  at low temperatures, and the results are shown in the main panel. The data were fitted with the exponential temperature dependence for a fully-open gap, $\Delta\lambda(T)~=~AT^{-\frac{1}{2}}{\rm e}^{-\varDelta(0)/k_BT}~+~B$, where $\varDelta(0)$ is the gap magnitude at zero temperature  and the constant $B$ allows for some variation in the extrapolated zero temperature value. The fitting was performed up to 0.12~K ($\approx T_c/5$) and as shown by the solid line in  Fig.~2(a), the model can account for the data with a gap of $\varDelta(0)~=~0.48k_BT_c$. A similar gap value of $\varDelta(0)~=~0.58k_BT_c$ is obtained from a corresponding fit for the $A/S$-type sample, as displayed in the main panel of  Fig.~2(b). The small gap values in both cases means that  $\Delta\lambda(T)$ only saturates at very low temperatures. The results indicate similar superconducting properties  of the $S-$ and $A/S$-type samples and are consistent with the fully gapped superconductivity reported for an $S$-type single crystal  in  Ref.~\cite{Kittaka2014}.

\begin{figure}[t]
\centering
\includegraphics[width=.8\linewidth]{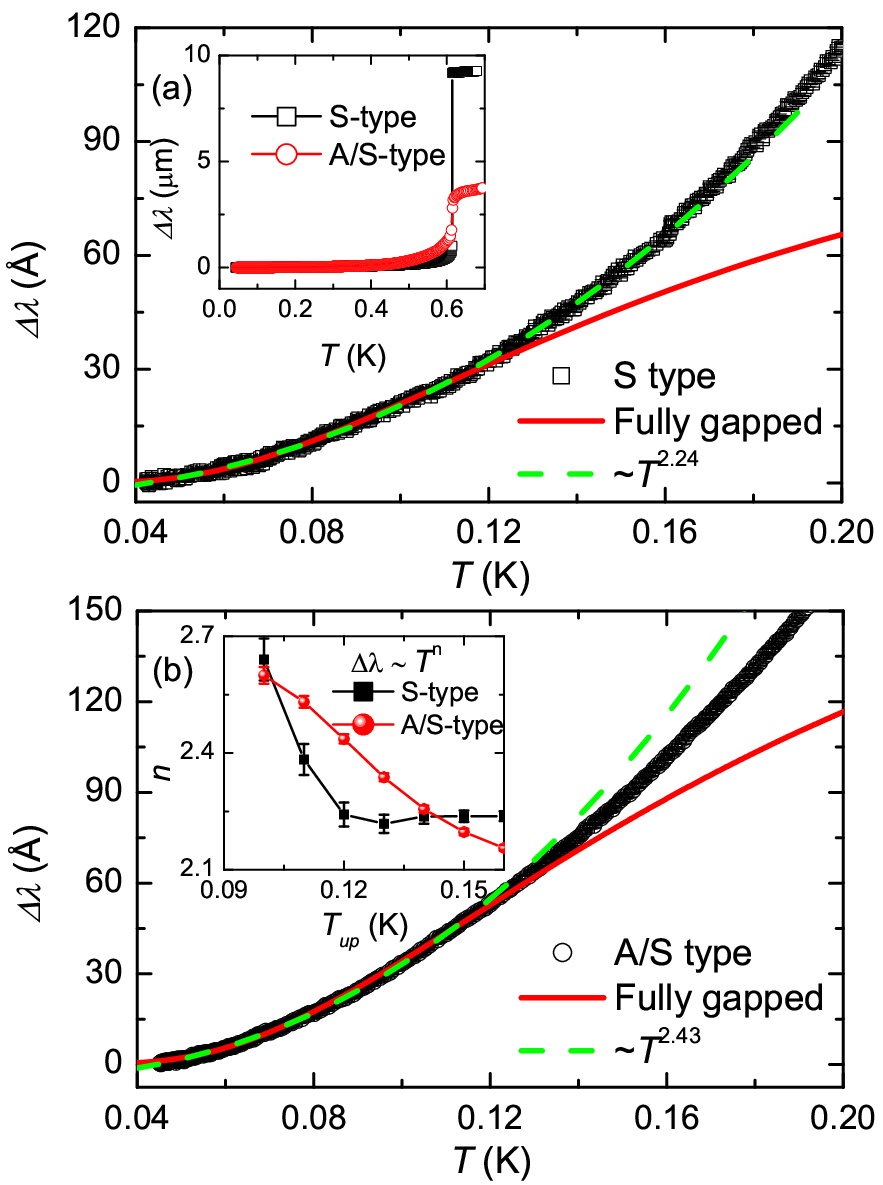}
\caption{The change in London penetration depth $\Delta\lambda(T)$ at low temperature for an (a) $S$-type and (b) $A/S$-type sample of CeCu$_2$Si$_2$. The solid lines show fits to a fully gapped model described in the text while the dashed lines show fits to a power law temperature dependence of $\Delta\lambda(T)\sim T^n$. The data across the whole temperature range of the superconducting states are displayed in the inset of (a). The inset of (b) shows $n$ when the data are fitted with $\Delta\lambda(T)\sim T^n$ up to a temperature $T_{up}$.}
\label{LamT}
\end{figure}

 The penetration depth of the $S$- and $A/S$-type samples could also be described by a power law dependence $\sim T^n$ ($SI Appendix$), with $n~=~2.24$ and 2.43 respectively, when fitting from the base temperature to 0.12~K. For line nodes in the superconducting gap in the presence of impurity scattering, $\Delta\lambda(T)$ may show quadratic behavior at low temperatures which crosses over to linear behavior at an elevated temperature \cite{SCImp}.  To check how the exponent $n$ evolves with temperature, we also fitted with the power law expression from the base temperature up to a range of temperatures $T_{up}$, and the dependence of $n$ on  $T_{up}$ is shown in the inset of Fig.~2(b). It can clearly be seen that for both samples, $n$ increases with decreasing  $T_{up}$, with $n~>~2$. This indicates that the true low temperature behavior is not a $\sim T^2$ dependence, as expected for a dirty nodal superconductor, but  $n$ increases as expected for  superconductivity exhibiting a full gap.  Therefore both the specific  heat and $\Delta\lambda(T)$ data are consistent with fully gapped superconductivity at very low temperatures.

\subsection*{Analysis of the superfluid density}

The superfluid density was calculated using $\rho_s(T)=[\lambda(0)/\lambda(T)]^2$  and is displayed for both $S$- and $A/S$-type samples in Fig.~3, where $\lambda(0)=2000$~\AA~\cite{CCSPen0}. The superfluid density was fitted following the method of  Ref.~\cite{pendepthrev}, for a gap $\varDelta_k(T)$ integrated over a cylindrical Fermi surface. The superfluid density data were fitted with an isotropic $s$-wave model with a gap $\varDelta(T)=\varDelta(0){\rm tanh}\left[1.82\left(1.018\left(\frac{T_c}{T}-1\right)\right)^{0.51}\right]$ \cite{SCgap}, as well as a $d$-wave model with line nodes  ($\varDelta_k(T,\phi)=\varDelta(T)\rm{cos}2\phi$, $\phi$=azimuthal angle). As shown in Fig.~3(a), a single-band isotropic  $s$-wave gap cannot account for the data of the $S$-type sample, in contrast to the low temperature $\Delta\lambda(T)$ data discussed above [Fig.~2(a)] . The 
single-band
$d$-wave model  shows reasonable agreement above $0.5~T_c$,   but the agreement is poor at low temperatures, since for this model $\rho_s(T)$ is linear, but the data are not. The agreement with the $d$-wave model at higher temperatures is consistent with the previously reported evidence for $d$-wave superconductivity. The data were also fitted using a two-gap $s$-wave model \cite{SCgap}, and  this gives reasonable agreement. The fitted gap values are  $\varDelta_1(0)=1.96k_BT_c$ and $\varDelta_2(0)=0.75k_BT_c$ , with a fraction for the larger gap of $x_1=0.74$. Both gap values are slightly larger than the ones obtained for the two-gap model  in Ref.~\cite{Kittaka2014}. Similarly, neither the 
$s$- nor $d$-wave 
single-band
models could describe the superfluid density of the $A/S$-type sample
at low temperatures
[Fig.~3(c)], but the data could be fitted using a two-gap model with   $\varDelta_1(0)=2.0k_BT_c$,  $\varDelta_2(0)=0.75k_BT_c$, and $x_1=0.74$. 

\begin{figure}[t]
\centering
 \includegraphics[width=0.99\columnwidth]{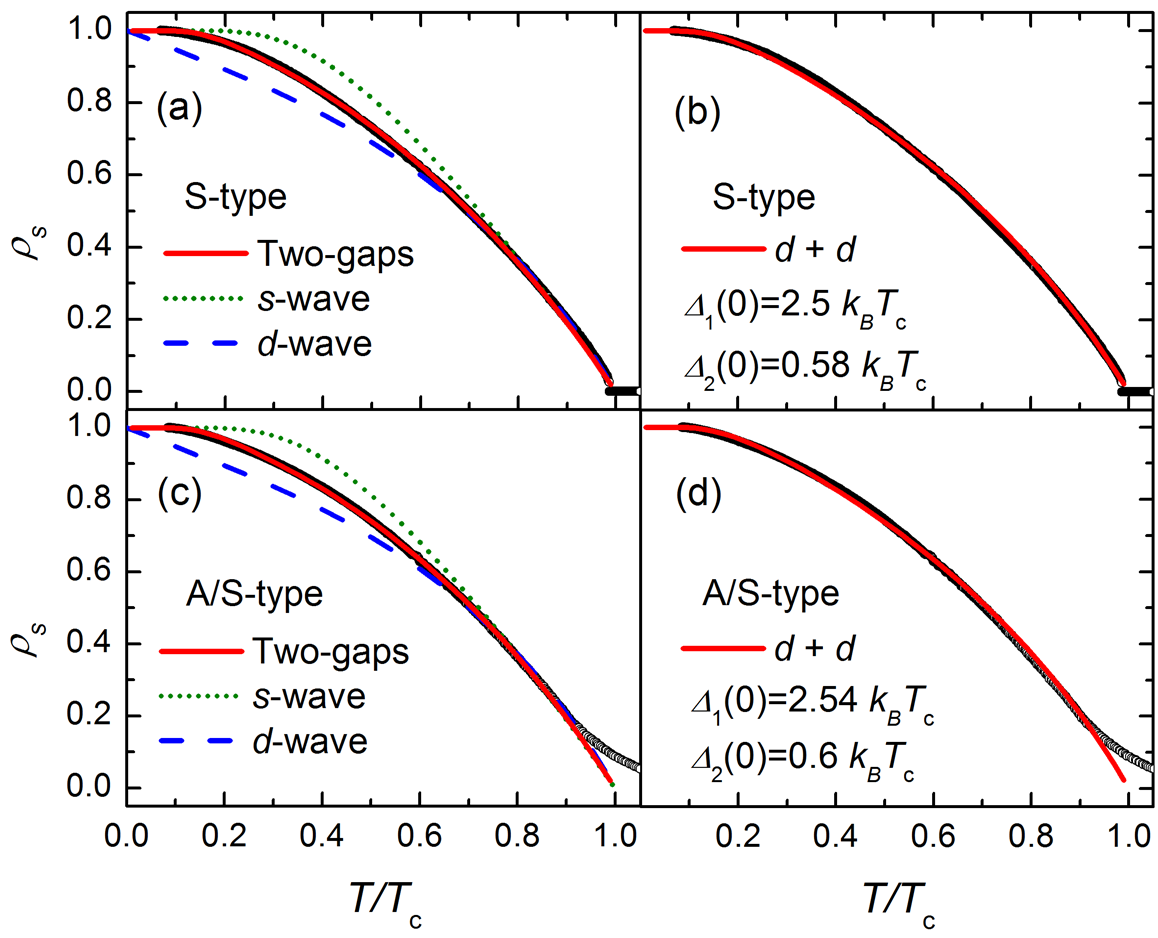}
\caption{Superfluid density of CeCu$_2$Si$_2$ fitted with various models. Fits for the $S$-type sample are shown for (a) two fully-open gaps, as well as $s$- and  $d$-wave models denoted by solid, dotted and dashed lines respectively, and (b) a  $d+d$ band-mixing pairing model. Fits for  the $A/S$-type sample are shown for (c) two fully-open gaps, as well as $s$- and  $d$-wave models denoted by solid, dotted and dashed lines respectively, and (d)  a $d+d$ band-mixing pairing model.}
\label{SuperF}
\end{figure}

On the other hand, it is difficult to reconcile an $s$-wave model with the evidence for a sign changing gap function, as concluded from the inelastic neutron scattering response, where a sharp spin resonance forms at the edge of a spin gap well inside the superconducting gap \cite{Stockert2011}. Moreover, the incommensurate ordering wave vector of the nearby SDW  (${\boldsymbol Q}_{\mathbf{AF}}$) is identical to the nesting wave vector spanning the flat parallel parts of the warped cylinders \cite{AphaseNeut}. This shows that there is a sign change of the pair wave function inside the dominating heavy-fermion band, which is incompatible with a nodeless $s_{\pm}$ pairing state \cite{Ikeda2015}. 
On the other hand, if 
there is no sign change of the gap function across the Fermi surface, $\Delta_{\mathbf k}\Delta_{\mathbf k+q}=|\Delta_{\mathbf k}||\Delta_{\mathbf k+q}|$, the coherence factor in the spin susceptibility $\chi''({\mathbf q},\omega)$ is vanishingly small \cite{SpinRes1,SpinRes2}. 
Consequently, in this case,
the spin spectrum will not have a sharp peak, although there may be a broad enhancement of the spectral weight \textit{above} $2\Delta$ \cite{SpinResTheorNo}. However,
when there is a change of sign of the gap function between regions of the Fermi surface connected by  ${{\boldsymbol q}}={\boldsymbol Q}_{\mathbf{AF}}$, there is an enhanced coherence factor since $\Delta_{\mathbf k}\Delta_{\mathbf k+q}=-|\Delta_{\mathbf k}||\Delta_{\mathbf k+q}|$. This gives rise to a sharp peak in $\chi''({\mathbf q},\omega) $\textit{ below} $2\Delta$, leading to the conclusion that there must be a sign changing order parameter in CeCu$_2$Si$_2$
\cite{SpinResTheor,SpinResTheor2}. 
By a similar argument, the lack of a coherence peak in NQR measurements also strongly disfavors superconductivity without a sign reversal  \cite{Ishidal1999,Fujiwara2008,NQR2017}. Furthermore,
given the strong Coulomb repulsion in Ce-based heavy fermion superconductors, the order parameter must be anisotropic with a sign change,
without running into the issue of a large $\mu^*$ \cite{Anderson-book}.
In other words, the $f$-electrons have a Coulomb repulsion that is much larger than their effective Fermi energy,
and they must avoid each other, thereby excluding any sign-preserving pairing function. In such strongly correlated superconductors, even anisotropic and sign-changing pairing states can be robust against 
disorder \cite{Anderson-book}. Indeed, potential scattering (due to a site exchange between Cu and Si of less than $1\%$ within the homogeneity range \cite{FSBook})  which enhances the residual resistivity by a factor of $\approx$4, apparently has an almost negligible influence on $T_c$, cf. the resistivity results on the $S$- and $A/S$-samples in Figs.~1(a) and (b). 
Also, recent  experiments on electron-irradiated samples revealed only a minor change of $T_c$ \cite{Yamashita2017}. At the same time, 
just like in the cuprates, the effect of substitutional disorder on $T_c$ is known to be site and size dependent \cite{Spille1983}.
For CeCu$_2$Si$_2$, the
superconducting $T_c$ was found to be extremely sensitive to non-magnetic substitutions
on the Cu site: for example, Rh, Pd, and Mn substitution for Cu at a level of $\approx$1 at$\%$ fully suppresses superconductivity \cite{Spille1983}, which is impossible to account for in the scenario of an $s$-wave state without a sign change of the order parameter. Further studies are needed to develop a detailed understanding of all these observations. 

\begin{figure}[t]
\centering
\includegraphics[width=.8\linewidth]{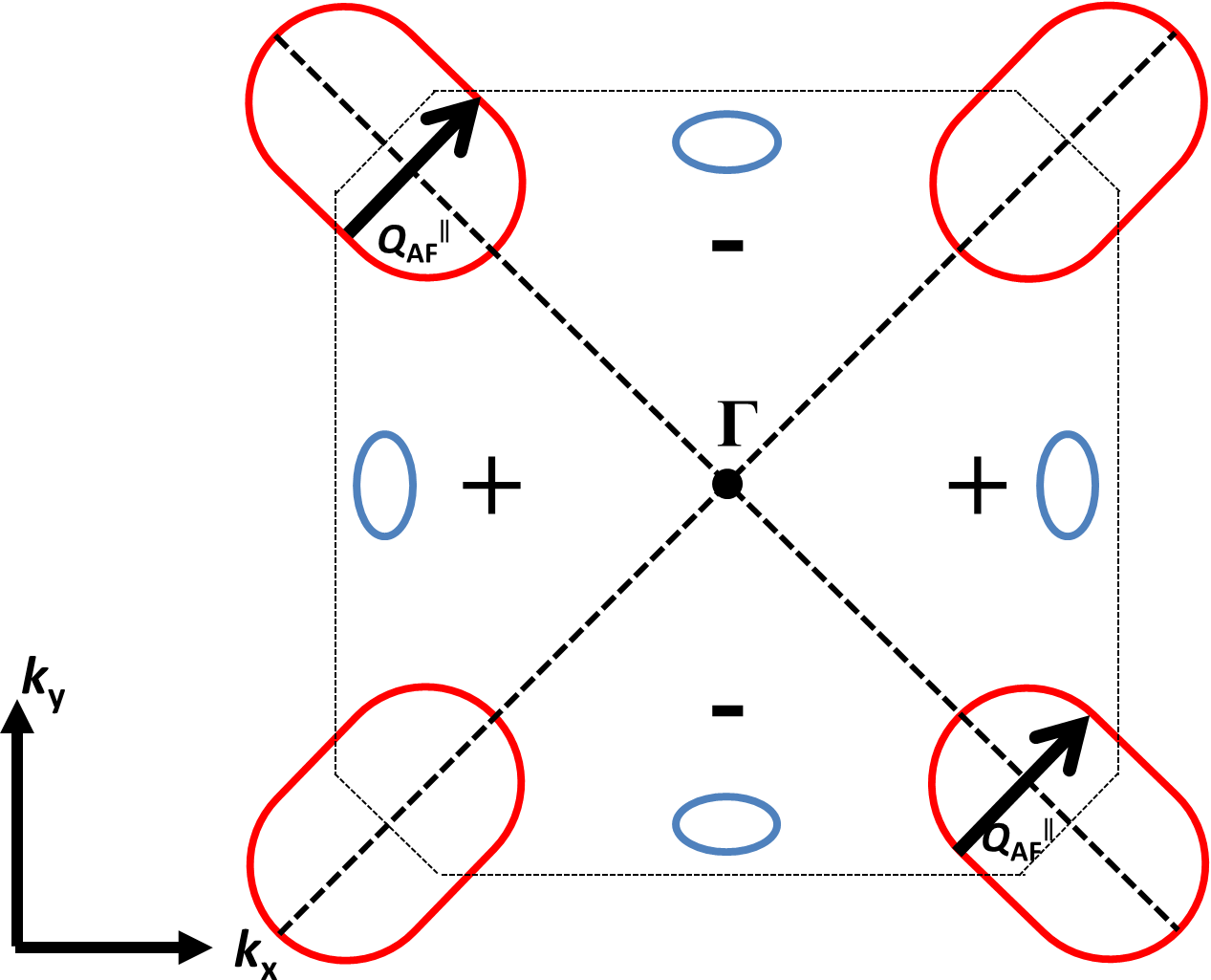}
\caption{An illustration of the warped parts of the cylindrical Fermi surfaces (red) in CeCu$_2$Si$_2$ at particular values of $k_z$, corresponding to the nesting portions of the three-dimensional Fermi surface, as well as additional smaller pockets (blue)  projected onto the $k_x - k_y$ wavevector plane \cite{CCSBS}. The component ${\boldsymbol Q}_{\mathbf{AF}}^\parallel$ of the antiferromagnetic wavevector ${\boldsymbol Q}_{\mathbf{AF}}$ projected into the same wavevector plane connects the parts of the heavy Fermi surface with a sign change in the intraband pairing component. The corresponding Fermi surface and nesting wavevector ${\boldsymbol \tau}~=~{\boldsymbol Q}_{\mathbf{AF}}$ in the three-dimensional space are those displayed in Fig. 3(b) of Ref.~ \cite{AphaseNeut}.}
\label{Fig4}
\end{figure}

In the present work, we consider a  pairing function which, by analogy with an $s \tau_3$ pairing state \cite{NicaTheor}, has an effective gap as a result of intraband pairing with $d_{x^2-y^2}$ symmetry and  interband pairing with $d_{xy}$ symmetry; our pairing function preserves both the fully gapped nature and order parameter sign change along ${\boldsymbol Q}_{\mathbf{AF}}$ 
on  a single nested Fermi surface. The $s \tau_3$ pairing state was introduced in the context of the iron-based superconductors ($SI Appendix$) \cite{NicaTheor}, as part of the studies about orbital-selective superconducting pairing \cite{Goswami10,Yu14,Ong13,Ong16}. There, the pairing function has the form $\varDelta \sim s_{x^2y^2}({\boldsymbol k}) \times \tau_3$ (``$s \tau_3$"), as a product of an $s$-wave form factor and a Pauli matrix in the $d_{xz}, d_{yz}$ orbital sub-space. For that case, the inter-orbital mixing in the dispersion part of the Hamiltonian ensures that in the \emph{band basis} the pairing is equivalent to a superposition of intra- and inter-band components with $d_{x^2-y^2}({\boldsymbol k})$ and $d_{xy}({\boldsymbol k})$ form factors respectively. The resulting quasiparticle spectrum acquires a non-vanishing $\left|\varDelta({\boldsymbol k}) \right|^2$ contribution, as the two components are added in quadrature, ensuring a full gap  on the whole Fermi surface with a sign change of the intraband component of the gap function. It is also shown how this pairing channel can be stabilized within a self-consistent five-orbital model, with a full gap and a resonance in the spin-excitation spectrum. Similar to the case considered in Ref.~\cite{NicaTheor}, ${\boldsymbol Q}_{\mathbf{AF}}$ of CeCu$_2$Si$_2$ will connect two parts of the Fermi surface with a sign change  in the intraband component of the gap function (see Fig.~4), thereby generating an enhanced spin spectral weight just above a threshold energy $E_0~=~ 3.9{\rm k_B}T_c$ \cite{Stockert2011},  inside the superconducting gap $2\Delta_1\approx5~k_BT_c$, see below and \cite{Fujiwara2008}.

In the following, we apply a simplified model for the gap structure to CeCu$_2$Si$_2$, given by summing contributions from the two $d$-wave states in quadrature, with $\varDelta(T,\phi)~= [(\varDelta_1(0)\rm{cos}(2\phi))^2+(\varDelta_2(0)\rm{sin}(2\phi))^2]^{\frac{1}{2}}\delta(T)$, where $\delta(T)$ is the gap temperature dependence  from Bardeen-Cooper-Schrieffer theory \cite{Bardeen1957}, which we used previously. In general, a $d+d$ band-mixing pairing can introduce corrections to the gap given above, due to the non-degeneracy of the  bands throughout the Brillouin zone, which would then lead to an extra parameter, the band splitting. In the following,
we will show that the data can be well fit by the simple function without this extra parameter. Although the $d_{xy}$ and $d_{x^2-y^2}$ states each have two line nodes, the nodes of the two states are offset by $\pi/4$ in the $k_x - k_y$~plane and as a result, the gap function is nodeless everywhere on a cylindrical Fermi surface. It should also be noted that for this model the same $\rho_s(T)$ is calculated upon exchanging $\varDelta_1(0)$ and $\varDelta_2(0)$. 
The superfluid density of the $S$-type sample was fitted using this model and the results are shown in Fig.~3(b). It can be seen that such a model can also fit the data well, with gap parameters of  $\varDelta_1(0)=2.5k_BT_c$ and $\varDelta_2(0)=0.58k_BT_c$. In Fig.~3(d), $\rho_s(T)$ for the $A/S$-type sample is equally well fitted, with similar parameters of  $\varDelta_1(0)=2.54k_BT_c$ and $\varDelta_2(0)=0.6k_BT_c$. The values of the larger gap agree almost perfectly with the gap value obtained from Cu-NQR measurements at higher temperatures \cite{Fujiwara2008}. Furthermore, this model only uses two fitting parameters, while the two band $s$-wave model of  Ref.~\cite{Kittaka2014} needs three.

\subsection*{Analysis of the temperature dependence of the specific heat}

\begin{figure}[t]
\centering
\includegraphics[width=0.8\columnwidth]{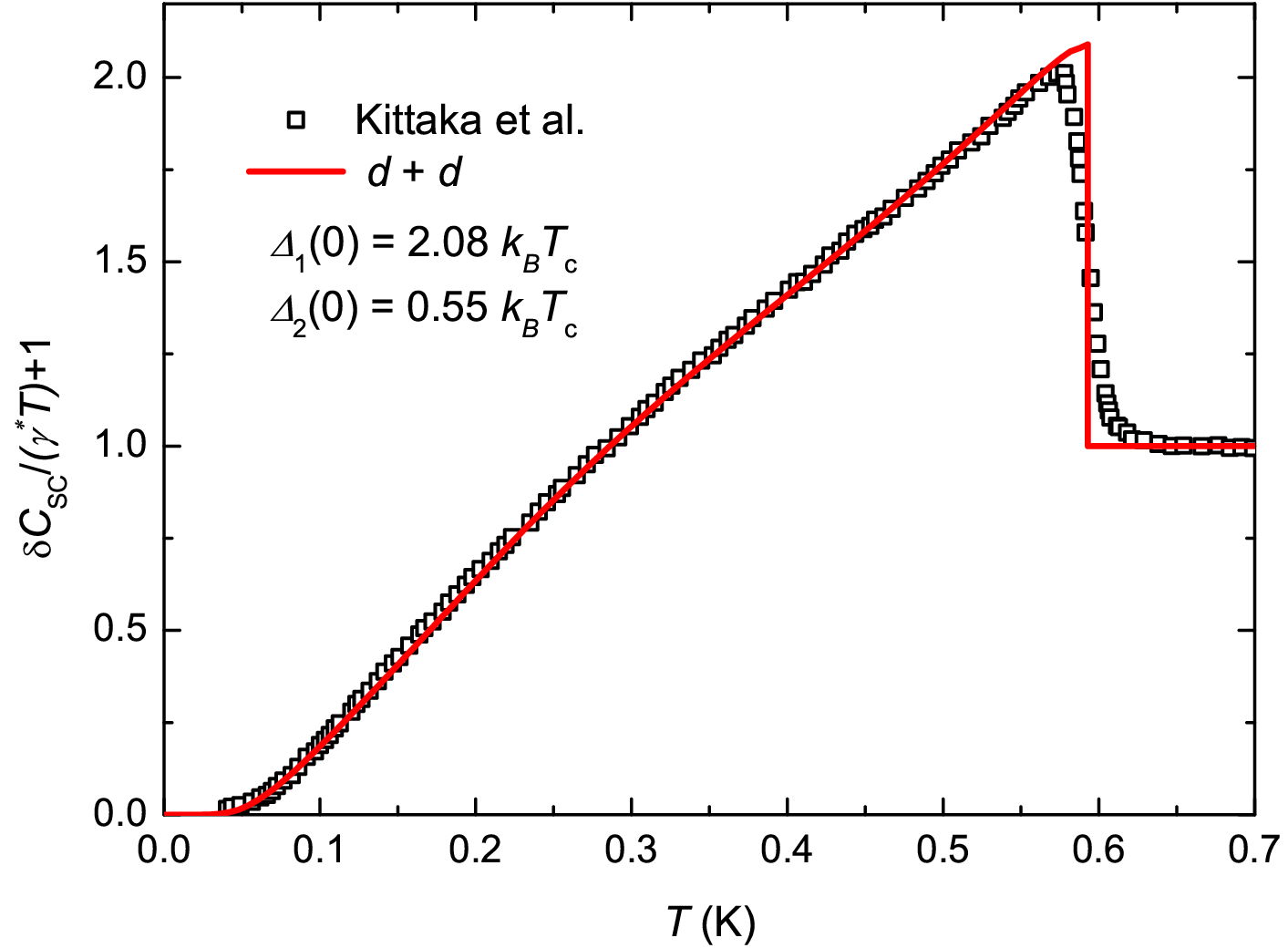}
\caption{Specific heat of $S$-type CeCu$_2$Si$_2$ digitized from Ref.~\cite{Kittaka2014}. The solid line shows a fit to the  $d+d$ band-mixing pairing model.}
\label{FigSpecH}
\end{figure}

We also reanalyzed the specific heat data digitized from Ref.~\cite{Kittaka2014} using the $d+d$ band-mixing pairing model. As shown in Fig.~5, the data can also be well described using this model,  with fitted parameters $\varDelta_1(0)=2.08~k_BT_c$ and $\varDelta_2(0)=0.55~k_BT_c$. The value of the small gap is similar to that obtained from the superfluid density fit, while the large gap is smaller in comparison. It should be noted that the calculated superfluid density requires an estimation of $G/\lambda(0)$, where $G$ is a calibration constant for the TDO method,  and the small differences in the gap values from the fits may arise due to uncertainties in this value.

\section*{Discussion and summary}

Both the superfluid density and specific heat results are highly consistent with a  model of the $d+d$ band-mixing pairing state, which 
most importantly
also explains the sign change of the superconducting order parameter. Although the fully gapped nature of the pairing state means that the density of states $N(E)$ is zero at low energies, $N(E)$ is nearly linear above the small gap, much like for pairing states with line nodes. This is also consistent with the literature results  showing $d$-wave superconductivity \cite{Ishidal1999,Fujiwara2008}, which were not obtained at low enough temperatures to observe clear evidence for fully gapped behavior. The lack of a coherence peak below $T_c$ in  the Cu-NQR $1/T_1(T)$ measurements \cite{Ishidal1999,Fujiwara2008,NQR2017} can also not be  accounted for by a  two-gap $s$-wave model, but is readily taken into account  by the anisotropic  $d+d$ state, which changes sign along  ${\boldsymbol Q}_{\mathbf{AF}}$ across the Fermi surface. We note that the effective gap corresponding to a $d+d$ band-mixing pairing is formally identical to one obtained from a $d+id$ pairing and the good fits to $\rho_s(T)$ and the specific heat are also consistent with this pairing.  In the $d+id$ state, time-reversal symmetry would be broken and although no clear experimental indication of a time-reversal symmetry breaking superconducting state was found from $\mu$SR measurements of $A/S$ type samples \cite{CCSMuSR}, this requires further study. By contrast, a $d+d$ band-mixing pairing is invariant under time-reversal, while generating the expected sign-change. 

From a theoretical perspective, 
unconventional superconductors in the presence of strong correlations are generally expected to be robust against 
disorder \cite{Anderson-book}. This has been demonstrated in models for strongly correlated superconductivity 
driven by short-range spin-exchange interactions\cite{Chkraborty17,Garg08}. 
The $s \tau_3$ pairing state \cite{NicaTheor} arises in a similar fashion, 
and is also expected to be robust against disorder. Because the $4f$-electrons in heavy fermion systems 
undoubtedly have strong correlations,
the $d+d$ band-mixing pairing state proposed for
CeCu$_2$Si$_2$ 
should be similarly robust to disorder, except for atomic substitutions \cite{Spille1983,Alloul2009}.

In conclusion, we have studied the change of penetration depth  $\Delta\lambda(T)$ and normalized superfluid density $\rho_s(T)$ of the heavy fermion superconductor CeCu$_2$Si$_2$ (both $A/S$- and $S$- type samples). The  behavior of  $\Delta\lambda(T)$ at very low temperatures  agrees with fully gapped superconductivity,  as concluded from specific heat measurements \cite{Kittaka2014}. We demonstrate that a nodeless  $d+d$ band-mixing pairing state can account for  the temperature dependence of both the superfluid density and specific heat. This state has the necessary sign change of the superconducting order parameter along ${\boldsymbol Q}_{\mathbf{AF}}$ on the heavy Fermi surface deduced from INS \cite{Stockert2011}, and is consistent with the lack of a coherence peak in $1/T_1(T)$. The model  also explains the consistency of $d$-wave superconductivity at higher temperatures, previously reported from  $1/T_1(T)$ measurements \cite{Ishidal1999,Fujiwara2008,NQR2017}. We therefore propose this $d+d$ band-mixing pairing state to be the superconducting order parameter of CeCu$_2$Si$_2$. Given that this is also a strong candidate pairing state for FeSe based superconductors \cite{NicaTheor}, such a pairing model may well be applicable to a wider range of fully gapped unconventional superconductors, including the case of a single CuO$_2$ layer \cite{CuO2mono}.

\section*{Methods}

CeCu$_2$Si$_2$ single crystals  were synthesized by a modified Bridgman technique using a self-flux method \cite{Seiro2010}. The temperature dependence of the London penetration depth shift $\Delta\lambda(T)=G\Delta f(T)$ was measured down to about 40~mK  using a tunnel diode oscillator (TDO) based technique \cite{TDOmethod}, where $\Delta f(T)~=~f(T)-f(0)$, $f(T)$ is the resonant frequency of the TDO coil, and $G$ is a calibration constant \cite{Gfactor}.

\section*{Acknowledgments}

We would like to thank R. Yu, S. Kirchner, G. Zwicknagl, P. Coleman, D. J. Scalapino, O. Stockert, S. Kittaka, and M. B. Salamon  for valuable discussions. The work at Zhejiang University has been supported by the National Key R\&D Program of China  (Grants No.~2017YFA0303100 and No.~2016YFA0300202), National Natural Science Foundation of China (U1632275, 1147425, and 11604291) and the Science Challenge Project of China (Project No.~TZ2016004). The work at Rice University has been supported by the NSF Grant No. 
DMR-1611392, the Robert A. Welch Foundation Grant No. C-1411, a QuantEmX grant from the Institute for Complex Adaptive Matter (ICAM) and the Gordon and Betty Moore Foundation through Grant No. GBMF5305 (Q.S.),
and the Center for Integrated Nanotechnologies, a U.S. DOE Basic Energy Sciences (BES) user facility. The work is also supported by the Sino-German Cooperation Group on Emergent Correlated Materials (GZ1123). Q.S. acknowledges the hospitality of the Aspen Center for Physics (NSF grant No. PHY-1607611) and the University of California, Berkeley. 

\section*{Author contributions}

The project was designed by  H.Q.Y.. The crystals were grown by H.S.J., Y.J.Z. W.X., H.L.,  and P.G..  G.M.P., M.S., J.L.Z., L.J., Z.F.W., Y.C., W.B.J. and H.Q.Y. performed the measurements, which were analyzed by G.M.P., M.S., F.S. and H.Q.Y. based on the theoretical model conceived by E.M.N. and Q.S..   The manuscript was written by G.M.P, M.S, F.S., J.L., Q.S. and H.Q.Y.

\clearpage
\renewcommand{\thefigure}{S\arabic{figure}}
\renewcommand{\thetable}{S\arabic{table}} 
\renewcommand{\theequation}{S\arabic{equation}} 
\setcounter{figure}{0}

\begin{center}
\noindent \textbf{{\LARGE Supporting information}} 
\end{center}

\subsection*{Theory of the $d+d$ pairing}
\emph{A priori}, the $d+d$ multiband pairing functions used in the analysis described in the main text can 
(i) produce a full gap along the Fermi pockets illustrated in Fig.~4 (main text) and
(ii) introduce a sign change between regions of the Fermi surface connected by 
the antiferromagnetic wavevector $\pmb{Q}_{AF}$ and hence allow for the presence of an in-gap (below $2\Delta$) peak
in the dynamical spin susceptibility.
Reconciling points (i) and (ii) is crucial for any proposed pairing, as indicated by the experiments 
discussed in the main text. 
This pairing function has clear conceptual advantages over 
the more conventional single and two $s$-wave gaps or the single $d$-wave gap, all of which are 
also minimal and phenomenological, but have difficulty allowing for \emph{both} requirements (i) and (ii).

A pairing of multiband $d+d$ type was initially proposed in Ref.~36
by two of the authors, 
in the context of similar 
experimental results for Fe-chalcogenide superconductors (SC).  In view of the availability of 
detailed effective tight-binding models and of well-established orbital-dependent correlations
in that case, the pairing was introduced in an orbital basis. Dubbed an ``$s\tau_{3}$"
pairing, it consisted of the direct product of an $s$-wave form factor and a Pauli matrix $\tau_{3}$
in the space of $3d_{xz}$ and $3d_{yz}$ orbitals, 
which altogether change sign under a $\pi/4$ rotation in the Brillouin zone (an irreducible 
 $B_{1g}$ representation 
of the $D_{4h}$ point group). 
One of the essential aspects was that in a well-defined, minimal
two-orbital model, the non-interacting band part and the pairing matrices do not commute. 
In an equivalent band basis, this results in \emph{both intra- and inter-band} pairing components, 
which are associated with $d_{x^{2}-y^{2}}(\pmb{k})$ and $d_{xy}(\pmb{k})$ 
form factors respectively:
\noindent \begin{equation}
\label{Eq:stau3}
{\Delta} (\pmb{k}) = \Delta_{1} (\pmb{k})  [d_{x^{2}-y^{2}}] \alpha_3 
+  \Delta_{2} (\pmb{k})   [d_{xy}] \alpha_1 .
\end{equation}
\noindent
where $\alpha_{3}$ and $\alpha_{1}$ are Pauli matrices in \emph{band}
space.
 The intra-band
pairing, denoted by $\alpha_{3}$,
must vanish along the diagonals of the BZ, while the inter-band pairing,
marked by $\alpha_{1}$,
 vanishes along the axes. 
Crucially, \emph{both components cannot vanish simultaneously}, excluding the zeroes of a common 
$s$-wave form factor. 
As an important consequence, the gap is essentially given by the quadrature of the two 
components and  is finite throughout  the FS. The situation in this case is illustrated by
 Fig.~1 in Ref.~36.

While the two-orbital model allows for analytical studies of the gap structure and excitation spectra, the $s\tau_{3}$
pairing is readily generalized to more complex cases. 
Indeed, in a more realistic five-orbital model for the Fe-chalcogenides, it was shown in Ref.~36
that 
the properties i) and ii) stated above are the robust characteristics of the pairing state.

While the analysis of a $d+d$ multiband pairing function in the case of the Fe-based SC
was greatly aided by the availability of established models, the essential insight into 
the advantages of minimal two-band $d+d$ pairing can be readily generalized to 
other SCs and to CeCu$_{2}$Si$_{2}$ in particular, on a phenomenological basis. 
 In this case, we propose a two-band minimal pairing matrix which parallels 
  \eqref{Eq:stau3} but simplifies 
$ \Delta_{1} (\pmb{k})  [d_{x^{2}-y^{2}}]$ and 
$\Delta_{2} (\pmb{k})  [d_{xy}]$ by 
$\Delta_{1} \cos(2\phi) $ and 
$\Delta_{2} \sin(2\phi)$ respectively:
\noindent \begin{equation}
\label{Eq:stau3-ccs}
\hat{\Delta} = \Delta_{1} \cos(2\phi) \alpha_{3} + \Delta_{2} \sin(2\phi) \alpha_{1},
\end{equation}

\noindent where the $\cos(2\phi)$ and $\sin(2\phi)$ stand for effective $d_{x^{2}-y^{2}}$ and 
$d_{xy}$ form factors respectively. The effective gap is then given by $|\hat{\Delta}|^{2}$, 
which, in view of the anti-commutation of the Pauli matrices, reproduces the expression
given in the main text. For  CeCu$_{2}$Si$_{2}$,
the analogue of Fig.~1 of Ref.~36 is given in Fig.~4 of the main text.

Finally, we note that $s\tau_3$ is one of the pairing states driven by magnetic fluctuations. Indeed, in the microscopic models for the Fe-based SCs in which unconventional superconductivity is induced for short-range spin-exchange interactions, the phase diagram for the superconducting pairing state was 
studied in Ref.~36. It was shown that the $s\tau_{3}$ pairing is stabilized 
over an extended region of the phase diagram. It is worth underscoring that the entire 
pairing function in 
 \eqref{Eq:stau3} --as opposed to the individual $ \Delta_{1} (\pmb{k})  [d_{x^{2}-y^{2}}]\alpha_3$ intraband or  $\Delta_{2} (\pmb{k})  [d_{xy}]\alpha_1$ interband component --
describes this pairing state.
This property carries over to the case of 
 \eqref{Eq:stau3-ccs} proposed for CeCu$_{2}$Si$_{2}$.  Thus, the entire pairing function corresponds to a single pairing channel and, in particular, the intraband and interband components admit no linear coupling and will have the same temperature dependence.

In short, the minimal $d+d$ multiband pairing provides a simple scenario which can in principle reconcile 
seemingly contradictory experimental signatures. In particular, it provides 
a good fit to the observed penetration depth and superfluid-density measurements. 

\subsection*{Fitting the low temperature penetration depth}

\begin{figure}[t]
\centering
\includegraphics[width=0.99\columnwidth]{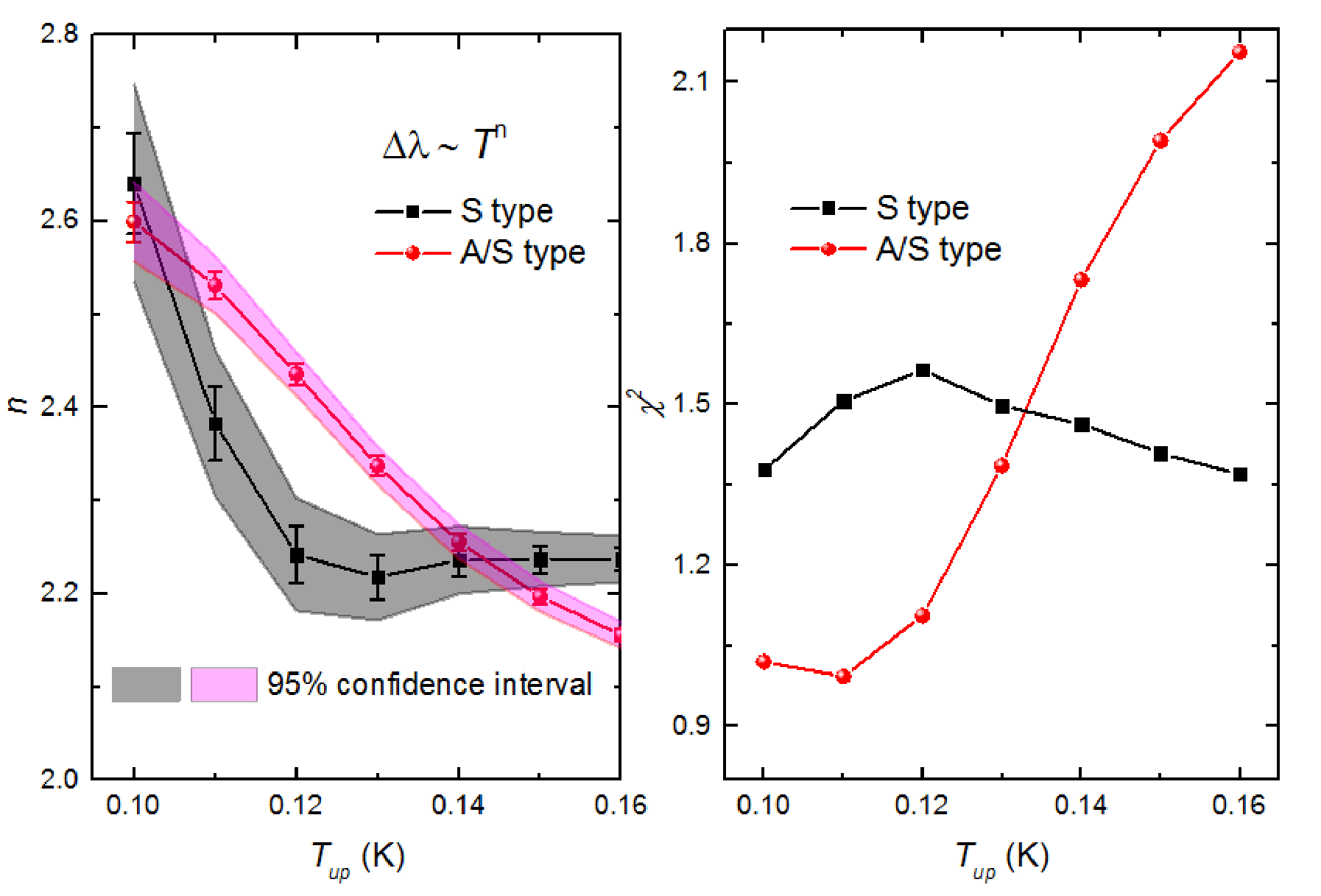}
\caption{The dependence of the exponent $n$ on  $T_{up}$ upon fitting $\Delta\lambda(T)$ with a  $\sim T^n$ dependence for both samples. The shaded areas show the area enclosed by the 95$\%$ confidence intervals. (b) The dependence of the goodness of fit $\chi^2$  on  $T_{up}$ for both samples.}
\label{FigErrSI}
\end{figure}

The temperature dependence of the London penetration depth shift $\Delta\lambda(T)=\lambda(T)-\lambda(0)$ was analyzed using both a model for a fully gapped superconductor and a power law relationship $\sim T^n$, as described in the main text. The fitting with the power law was performed from the base temperature of 40~mK up to a temperature $T_{up}$, and the dependence of the exponent $n$ on $T_{up}$ was obtained. The data were fitted using a standard weighted least squares method. The weights were given by $1/\sigma_i^2$, where $\sigma_i$ is the uncertainty of the $i^{th}$ data point. Since a large number of datapoints were measured for TDO, $\sigma_i$ were estimated from the standard deviation of the data within narrow temperature intervals.

The dependence of the exponent $n$ on  $T_{up}$ is displayed in Fig.~S1(a).  For both samples there is an increase of $n$ with decreasing temperature, with an exponent significantly larger than two. The shaded areas illustrates the 95$\%$ confidence intervals, showing that there is a high level of confidence that the exponent is greater than two at low temperatures. This is clear evidence for nodeless superconductivity. The goodness of fit, 
\noindent \begin{equation}
\label{GFit}
\chi^2=\frac{1}{N}\sum\limits_{i=1}^N (\lambda_i-\lambda_{fit})^2/\sigma_i^2,
\end{equation}

\noindent for the fits of the power law behavior is displayed in Fig.~S1(b), where $\lambda_i$ are the experimental datapoints,  $\lambda_{fit}$ are the calculated model values and $N$ is the number of degrees of freedom ($\approx$ number of datapoints). At low temperatures $\chi^2$ is close to one, indicating the data are well described by an exponent larger than two, but there is no evidence that the model overfits the data.

\end{document}